\documentclass{article}
\usepackage{graphicx} 
\usepackage{amsmath} 	
\usepackage{amsfonts} 

\title{Ising Hamiltonian for Coupled Fibre-Based Optical Parametric Oscillators}
\author{Marina Zajnulina \\
zajnulina@multitel.be}  
\date{November 2023}

\begin{document}

\maketitle

\begin{abstract}
Here, I derive an Ising Hamiltonian for $N$ coupled fibre-based Optical Parametric Oscillators (OPOs). For this, I use the Lugiato-Lefever equation. The derivation of the Hamiltonian does not pretend being mathematically correct, although it works under the assumption of plane waves propagating through the OPOs. Its objective is rather to be as close as possible to the Hamilton formalism than to be mathematically rigorous. The approach is based on the assumptions that are justified from the physical point of view and are implemented experimentally. The resulting Ising Hamiltonian contains all necessary terms that are connected to the appearance and evolution of phase solitons. 
\end{abstract}

\section{Start}
In this section, we derive an Ising Hamiltonian for $N$ coupled optical resonators as they are given by a fibre-based OPO cavity equipped with $N-1$ delay lines. The evolution of the complex amplitude of an optical field $u(T, \tau)$ within one OPO is well described by the following model [1]:
\begin{equation}
\frac{\partial u}{\partial T} = i|u|^{2}u + i\frac{\partial^{2}}{\partial\tau^{2}}u - (1+i\Delta)u + \mu u^{*},
\label{equ:engl}
\end{equation}
where $T$ denotes the "long" time that corresponds to the resonator roundtrip number and $\tau$ is the "short" time of the field evolution within one resonator roundtrip. In Eq.~\ref{equ:engl}, the fist expression on the right side denotes the optical Kerr effect, the second one the group-velocity dispersion (GVD) in fibres, the third expression describes the dissipative effects in the cavity given by linear optical losses and the cavity detuning $\Delta,$ and the last one denotes a coherent pump with coupling parameter $\mu.$

Let us now consider $N$ coupled resonators of the form as described in Eq.~\ref{equ:engl}. Assume diffractive, equal and symmetric coupling between these resonators. At least theoretically, the assumption of diffractive coupling is justified as the energy of the pump would spread across all resonators with increasing $T$ if only one of them is driven and each resonator couples to its neighbour (cf. [2, 4]). With enough effort, equal and symmetric coupling can be realised experimentally. With these assumptions, Eq.~\ref{equ:engl} takes the form:
\begin{equation}
\frac{\partial u_{i}}{\partial T} = i|u_{i}|^{2}u_{i} + i\frac{\partial^{2}}{\partial\tau^{2}}u_{i} - (1+i\Delta)u_{i} + \mu_{i} u^{*} + \hat{C}\left(u_{i+1} + u_{i-1} - 2u_{i}\right).
\label{equ:engl_Nres}
\end{equation}
Here, $i$ denotes the index of the $i-$th resonator, $i\in \{i,...,N\},$ and $\mu_{i}$ expresses the fraction of the pump that drives the $i-$th resonator, $\mu = \sum_{i=1}^{N} \mu_{i}.$ In general terms, $\hat{C}$ denotes the complex coupling between the resonators, $\hat{C} = (\hat{c}_{1} + \hat{c}_{2}),$ $\hat{c}_{1}, \hat{c}_{2}\in \mathbb{R}.$  

\section{Discrete Nonlinear Schr\"odinger Equation}
\label{sec:DNLS_full}
Here, we simplify Eq.~\ref{equ:engl_Nres}. For this, let us make and justify the following assumptions.
\begin{itemize}
\item The dispersion of the resonators is negligible. It can be achieved either by dispersion compensating techniques or by designing of the dispersion profile of the fibres involved. Assuming zero-GVD allows us to have only one time derivative. 
\begin{equation}
    \frac{\partial^{2}u_{i}}{\partial\tau^{2}} \rightarrow 0 \Longrightarrow \frac{\partial u_{i}}{\partial T} \rightarrow \frac{d u_{i}}{d T}.
\end{equation}
\item The pump compensates for the optical losses and the impact of the cavity detuning $\Delta.$ 
\begin{equation}
    -(1+i\Delta)u_{i} + \mu_{i}u^{*} = 0.
\end{equation}
\item For simplicity's sake, let's also assume a slow evolution of the light intensity inside the resonators, such that $|u_{i}(T)|^{2}\approx|u_{0/i}|^{2}$ where $u_{0/i}$ is time-independent initial amplitude. This assumption is indeed valid as the temporal changes of the field happen over several hundreds of roundtrips, i.e. for high values of $T.$
\item Further, we assume only weak coupling between the resonators, i.e. $|\hat{C}|\ll1.$ This assumption is realisable from the experimental point of view.
\item The pump amplitude $\mu$ is small. The pump serves only to compensate the cavity losses such that the OPO is prevented from lasing (Ref. [1]).
\end{itemize}
Thus, we have: 
\begin{equation}
    \frac{d u_{i}}{dT} = i|u_{0/i}|^{2}u_{i} + \hat{C}\left(u_{i+1} + u_{i-1} - 2u_{i}\right)
\end{equation}
or multiplied by complex $i$ and rearranged:
\begin{equation}
    i\frac{d u_{i}}{dT} + |u_{0/i}|^{2}u_{i} - i\hat{C}\left(u_{i+1} + u_{i-1} - 2u_{i}\right) = 0.
\end{equation}
The complex $i$ can be integrated in the coupling parameter by setting $c_{1} = 0$ in $\hat{C} =\left(c_{1}+ ic_{2}\right)$ and renaming $c_{2} =: C:$
\begin{equation}
    i\frac{d u_{i}}{dT} + |u_{0/i}|^{2}u_{i} + C\left(u_{i+1} + u_{i-1} - 2u_{i}\right) = 0.
    \label{equ:DNLS}
\end{equation}
This is a discrete nonlinear Schr\"odinger equation. Its complex conjugate reads as  
\begin{equation}
    -i\frac{d u_{i}^{*}}{dT} + |u_{0/i}|^{2}u_{i}^{*} + C\left(u_{i+1}^{*} + u_{i-1}^{*} - 2u_{i}^{*}\right) = 0.
    \label{equ:DNLS_cc}
\end{equation}
The easiest way to work with Eq.~\ref{equ:DNLS} and its complex conjugate is to use the plane wave ansatz:
\begin{equation}
    u_{n}(T) = u_{0/n}e^{i\left(\omega T + \kappa n\right)}
\end{equation}
with a real-valued amplitude $u_{0/n},$ frequency $\omega,$ and wave number $\kappa.$ Here, I changed the index resonator index $i$ to $n$ to avoid confusion between the inted and the complex $i.$ However, I'll proceed using $i$ as an index in the further course and switch to $n$ if it deems necessary. 

\section{Basic Hamiltonian}
We now will try to derive an Ising Hamiltonian for Eq.~\ref{equ:engl_Nres}. For this, we follow the procedure described in Ref. [3].
First, multiply Eq.~\ref{equ:DNLS} by $u_{i}^{*}$ and Eq.~\ref{equ:DNLS_cc} by $-u_{i}$ and add the resulting expressions. The result reads as
\begin{equation}
    i\frac{d}{dT}|u_{i}|^{2} - C\left(u_{i}\left[u_{i+1}^{*} + u_{i-1}^{*}\right] -\left[u_{i+1} + u_{i-1}\right]u_{i}^{*} \right)=0.
    \label{equ:conserve}
\end{equation}
Eq.~\ref{equ:conserve} has the form of a conservation law (continuity equation) with 
\begin{equation}
    Q: = \sum_{i=1}^{N} |u_{i}|^{2} = const
\end{equation}
having the meaning of the total energy and
\begin{equation}
    M := \sum_{i=1}^{N}\left(u_{i}\left[u_{i+1}^{*} + u_{i-1}^{*}\right] -\left[u_{i+1} + u_{i-1}\right]u_{i}^{*} \right) = const = 0
\end{equation}
describing the total momentum of the system.

To obtain a Hamiltonian, multiply Eq.~\ref{equ:DNLS} by $\left(u_{i+1}^{*} + u_{i-1}^{*} - 2u_{i}^{*}\right)$ and Eq.~\ref{equ:DNLS_cc} by $-\left(u_{i+1} + u_{i-1} - 2u_{i}\right)$ and add the resulting expressions. This step yields:
\begin{equation}
\begin{split}
    i&\left(\frac{d u_{i}}{d T} \left(u_{i+1}^{*} + u_{i-1}^{*} - 2u_{i}^{*}\right) + \left(u_{i+1} + u_{i-1} - 2u_{i}\right)\frac{d u_{i}^{*}}{d T}\right) +\\
    &+ |u_{0/i}|^{2}\left(u_{i}\left[u_{i+1}^{*} + u_{i-1}^{*}\right] - \left[u_{i+1} + u_{i-1}\right]u_{i}^{*}\right) =\\
    &=-i2\frac{d }{d T}|u_{i}|^2 + i\left(\frac{d u_{i}}{d T} \left[u_{i+1}^{*} + u_{i-1}^{*} \right] + \left[u_{i+1} + u_{i-1} \right]\frac{d u_{i}^{*}}{d T}\right) + \\
    & + |u_{0/i}|^{2}\left(u_{i}\left[u_{i+1}^{*} + u_{i-1}^{*}\right] - \left[u_{i+1} + u_{i-1}\right]u_{i}^{*}\right) = 0.
\end{split}
\label{equ:tweve}
\end{equation}
Now, multiply Eq.~\ref{equ:conserve} by $|u_{0/i}|^{2}$ and Eq.~\ref{equ:tweve} by $C$ and add the resulting expressions. It yields:
\begin{equation}
\begin{split}
    -i2C\frac{d}{d T}|u_{i}|^2 &+ i C\left(\frac{d u_{i}}{d T}\left[u_{i+1}^{*} + u_{i-1}^{*} \right] + \left[u_{i+1} + u_{i-1} \right]\frac{d u_{i}^{*}}{d T}\right) + \\
     &+ i|u_{0/i}|^2\frac{d}{dT}|u_{i}|^{2} =0.
\end{split}
\label{equ:thirteen}
\end{equation}
Eq.~\ref{equ:thirteen} has the form of the conservation law 
\begin{equation}
    \frac{dH_B}{dT}+ C\left(\frac{d u_{i}}{d T}\left[u_{i+1}^{*} + u_{i-1}^{*} \right] + \left[u_{i+1} + u_{i-1} \right]\frac{d u_{i}^{*}}{d T}\right) = 0
\end{equation}
with basic Hamiltonian defined as
\begin{equation}
    H_{B} := \left(|u_{0/i}|^{2} - 2C\right)|u_{i}|^2.
    \label{equ:sixteen}
\end{equation}
 
The Ising Hamiltonian is obtained when we integrate also the coupling term in Eq.~\ref{equ:thirteen}. With plane-wave ansatz, we get:
\begin{equation}
\begin{split}
    &\left(\frac{d u_{i}}{d T}\left[u_{i+1}^{*} + u_{i-1}^{*} \right] + \left[u_{i+1} + u_{i-1} \right]\frac{d u_{i}^{*}}{d T}\right)= \\
    &\left(e^{i\kappa}+ e^{-i\kappa}\right)\left(\frac{d u_{i}}{d T}u_{i}^{*} + u_{i}\frac{d u_{i}^{*}}{d T}\right)= \left(e^{i\kappa}+ e^{-i\kappa}\right)\frac{d}{dT}|u_{i}|^{2}.
\end{split}
\end{equation}
This expression can be integrated over time $T$ yielding:
\begin{equation}
    \left(e^{i\kappa}+ e^{-i\kappa}\right)|u_{i}|^{2}= \left(e^{i\kappa}+ e^{-i\kappa}\right)u_{i}u_{i}^{*}=[u_{i+1} + u_{i-1}]u_{i}^{*}.
    \label{equ:coupling_term}
\end{equation}
Thus, we have:
\begin{equation}
    H_{I} := |u_{0/i}|^{2}|u_{i}|^{2} - 2C|u_{i}|^{2} +C[u_{i+1} + u_{i-1}]u_{i}^{*} = E = const,
    \label{equ:Ising_first}
\end{equation}
where the first term accounts for the Kerr effect, the second one has the meaning of losses into the neighbouring resonators, and the third on denotes the coupling between the neighbouring resonators.

\section{Extension of the Basic Hamiltonian}
Let us now drop the assumption $-(1+i\Delta)u_{i} + \mu_{i}u^{*} = 0$ and try to integrate the contribution of the optical losses, detuning $\Delta,$ and the pump. For it, we repeat the procedure described in the previous section.

\subsection{Optical Losses and Cavity Detuning}
For simplicity's sake, consider only the part of Eq.~\ref{equ:engl_Nres} that relates to the optical losses and the cavity detuning: 
\begin{equation}
    \frac{d}{dT} u_{i} = -(1 + i\Delta) u_{i}.
\end{equation}
Now, multiply this equation by complex $i$ and get its complex conjugate form:
\begin{equation}
    i\frac{d}{dT} u_{i} + i(1 + i\Delta)u_{i} = 0
    \label{equ:cav}
\end{equation}
and, accordingly,
\begin{equation}
    -i\frac{d}{dT} u_{i}^{*} - i(1 - i\Delta)u_{i}^{*} = 0.
    \label{equ:cavcc}
\end{equation}
As previously, we multiply the first equation by $u_{i}^{*}$ and the second one by $-u_{i}.$ Then, we add the expressions and get: 
\begin{equation}
    i\frac{d}{dT}|u_{i}|^2 +i2|u_{i}|^{2} = 0, 
\end{equation}
which updates our continuity equations (Eq.~\ref{equ:conserve}) to the following form:
\begin{equation}
    i\frac{d}{dT}|u_{i}|^{2} - C\left(u_{i}\left[u_{i+1}^{*} + u_{i-1}^{*}\right] -\left[u_{i+1} + u_{i-1}\right]u_{i}^{*} \right) +i2|u_{i}|^{2}=0.
\end{equation}

This is an interesting result that shows that the cavity detuning implemented in Eq.~\ref{equ:engl} has no impact on the continuity equation Eq.~\ref{equ:conserve}. The losses, however, open the system for dissipation of the energy. They, thus, appear in the continuity equation as the last term on the left side.

To see whether optical losses and the cavity detuning contribute to the Hamiltonian, multiply Eq.~\ref{equ:cav} by $\left(u_{i+1}^{*} + u_{i-1}^{*} - 2u_{i-1}^{*}\right)$ and Eq.~\ref{equ:cavcc} by\newline
$-\left(u_{i+1} + u_{i-1} - 2u_{i-1}\right)$ and add the expressions: 
\begin{equation}
\begin{split}
    i&\left(\frac{d u_{i}}{d T} \left(u_{i+1}^{*} + u_{i-1}^{*} - 2u_{i}^{*}\right) + \left(u_{i+1} + u_{i-1}-2u_{i}\right)\frac{d u_{i}^{*}}{d T}\right) +\\
    + i&\left(u_{i} \left[u_{i+1}^{*} + u_{i-1}^{*} \right] + \left[u_{i+1} + u_{i-1}\right]u_{i}^{*}\right)+\\
    & -\Delta\left(u_{i}\left[u_{i+1}^{*} + u_{i-1}^{*}\right] - \left[u_{i+1} + u_{i-1}\right]u_{i}^{*}\right)-4i|u_{i}|^{2} = 0.
\end{split}
\label{equ:twenty}
\end{equation}
In fact, optical losses and the cavity detuning contribute to the Hamiltonian. With this result, Eq.~\ref{equ:tweve} generalises to

\begin{equation}
\begin{split}
    -i2\frac{d }{d T}|u_{i}|^2 &+ i\left(\frac{d u_{i}}{d T} \left[u_{i+1}^{*} + u_{i-1}^{*} \right] + \left[u_{i+1} + u_{i-1} \right]\frac{d u_{i}^{*}}{d T}\right) + \\
    & + i\left(u_{i} \left[u_{i+1}^{*} + u_{i-1}^{*} \right] + \left[u_{i+1} + u_{i-1}\right]u_{i}^{*}\right)+\\
    & + \left(|u_{0/i}|^{2} -\Delta\right)\left(u_{i}\left[u_{i+1}^{*} + u_{i-1}^{*}\right] - \left[u_{i+1} + u_{i-1}\right]u_{i}^{*}\right) -4i|u_{i}|^{2}= 0.
\end{split}
\label{equ:twentyone}
\end{equation}
To include the impact of the losses and cavity detuning into the Hamiltonian, we multiply now Eq.~\ref{equ:conserve} by $\left(|u_{0/i}|^{2} -\Delta\right)$ and Eq.~\ref{equ:twentyone} by $C$ and, again, add the expressions. Thus, we get an update for the conservation law:
\begin{equation}
\begin{split}
    &i\left(|u_{0/i}|^{2}-2C-\Delta\right)\frac{d}{d T}|u_{i}|^2 + 2i\left(|u_{0/i}|^{2}-2C-\Delta\right)|u_{i}|^{2}+\\
     &+ i C\left(\frac{d u_{i}}{d T}\left[u_{i+1}^{*} + u_{i-1}^{*} \right] + \left[u_{i+1} + u_{i-1} \right]\frac{d u_{i}^{*}}{d T}\right) + \\
     & + iC\left(u_{i} \left[u_{i+1}^{*} + u_{i-1}^{*} \right] + \left[u_{i+1} + u_{i-1}\right]u_{i}^{*}\right)=0.
\end{split}
\label{equ:twentytwo}
\end{equation}
Remember, the integration of this expression over time $T$ will deliver us the Ising Hamiltonian.

\subsection{Contribution of the Pump}
Consider now
\begin{equation}
    i\frac{d}{dT}u_{i} = i\mu_{i}u^{*}\quad \text{and} \quad -i\frac{d}{dT}u_{i}^{*} = -i\mu_{i}u.
\end{equation}
For the contribution to the continuity equation Eq.~\ref{equ:conserve}, we get:
\begin{equation}
    i\frac{d}{dT}|u_{i}|^{2} - i\mu_{i}\left(u u_{i} + u^{*}u_{i}^{*} \right)=0.
\end{equation}
The coupling between the pump and the resonator field generates an oscillation at twice the frequency which can be easily seen with a plane-wave ansatz.

The continuity equation Eq.~\ref{equ:conserve} generalises to 

\begin{equation}
\begin{split}
    i\frac{d}{dT}|u_{i}|^{2} &- C\left(u_{i}\left[u_{i+1}^{*} + u_{i-1}^{*}\right] -\left[u_{i+1} + u_{i-1}\right]u_{i}^{*} \right) +\\  
    &+2i|u_{i}|^{2}- i\mu_{i}\left(u^{*}u_{i}^{*} + u u_{i}\right) =0.
    \end{split}
    \label{equ:conserve_pump}
\end{equation}

And Eq.~\ref{equ:twentyone} takes the form:
\begin{equation}
\begin{split}
    -i2\frac{d }{d T}|u_{i}|^2 &+ i\left(\frac{d u_{i}}{d T} \left[u_{i+1}^{*} + u_{i-1}^{*} \right] + \left[u_{i+1} + u_{i-1} \right]\frac{d u_{i}^{*}}{d T}\right) + \\
    & + i(1-\mu_{i})\left(u_{i} \left[u_{i+1}^{*} + u_{i-1}^{*} \right] + \left[u_{i+1} + u_{i-1}\right]u_{i}^{*}\right)+\\
    & + \left(|u_{0/i}|^{2} -\Delta\right)\left(u_{i}\left[u_{i+1}^{*} + u_{i-1}^{*}\right] - \left[u_{i+1} + u_{i-1}\right]u_{i}^{*}\right) -\\
    & -4i|u_{i}|^{2}+ 2i\mu_{i}[u_{i}u + u^{*}u_{i}^{*}]= 0.
\end{split}
\label{equ:twentynine}
\end{equation}

Now, again, we multiply the continuity equation Eq.~\ref{equ:conserve_pump} by $\left(|u_{0/i}|^{2} - \Delta\right)$ and Eq.~\ref{equ:twentynine} by $C$ and add the expressions. We get:
\begin{equation}
\begin{split}
    &i\left(|u_{0/i}|^{2}-2C-\Delta\right)\frac{d}{d T}|u_{i}|^2 + 2i\left(|u_{0/i}|^{2}-2C-\Delta\right)|u_{i}|^{2}-\\
    &-i\mu_{i}\left(|u_{0/i}|^{2}-2C-\Delta\right)\left(u_{i}u + u^{*}u_{i}^{*}\right)+\\
     &+ i C\left(\frac{d u_{i}}{d T}\left[u_{i+1}^{*} + u_{i-1}^{*} \right] + \left[u_{i+1} + u_{i-1} \right]\frac{d u_{i}^{*}}{d T}\right) + \\
     & + iC\left(u_{i} \left[u_{i+1}^{*} + u_{i-1}^{*} \right] + \left[u_{i+1} + u_{i-1}\right]u_{i}^{*}\right)-\\
     &- i\mu_{i}C\left(u \left[u_{i+1}^{*} + u_{i-1}^{*} \right] + \left[u_{i+1} + u_{i-1}\right]u^{*}\right)=0.
\end{split}
\label{equ:thirty}
\end{equation}
Eq.~\ref{equ:thirty} contains all necessary terms now that are needed for the final Ising Hamiltonian. The latter is obtained by integration of Eq.~\ref{equ:thirty} over time $T.$ 

\section{Simplify!}
Eq.~\ref{equ:thirty} is a bit difficult to interpret (and also quite difficult to integrate), although it already contains all necessary terms and relations of an Ising Hamiltonian. Let's see whether we can find a slightly simpler and easier to understand form. Eq.~\ref{equ:engl} is normalised. Let's equip it with factors that are usually used in nonlinear optics. Later, it will help us see the meaning of the Hamiltonian terms.
\begin{equation}
\frac{\partial u}{\partial T} = i\gamma|u|^{2}u + i\beta_{2}\frac{\partial^{2}}{\partial\tau^{2}}u - (\frac{\alpha}{2}+i\Delta)u + \mu u^{*},
\label{equ:engl_fac}
\end{equation}
where we introduced $\gamma$ as the nonlinear parameter that describes the Kerr effect, $\beta_{2}$ is the group-velocity dispersion (GVD) parameter, and $\alpha$ describes the nonlinear losses. All other parameters ($\Delta,$ $\mu,$ and in the next step $C$) have the same meaning as previously although their units are different now.

Let's assume (non-diffractive) next-neighbour coupling. With all other assumptions we made in Sec.~\ref{sec:DNLS_full}, we get 
\begin{equation}
    i\frac{d u_{i}}{dT} + \gamma|u_{0/i}|^{2}u_{i} + C\left(u_{i+1} + u_{i-1}\right) = 0.
    \label{equ:DNLS_sh}
\end{equation}
and 
\begin{equation}
    -i\frac{d u_{i}^{*}}{dT} + \gamma|u_{0/i}|^{2}u_{i}^{*} + C\left(u_{i+1}^{*} + u_{i-1}^{*}\right) = 0.
    \label{equ:DNLS_cc_sh}
\end{equation}
Although less obvious, these equations are a type of  discrete Nonlinear Schr\"odinger equations. They belong to the class of Discrete Self-Trapping (DST) equations ([5]):
\begin{equation}
    i\frac{d u_{i}}{dT} + \gamma|u_{0/i}|^{2}u_{i} + C\sum_{k\neq i}^{N}u_{k} = 0.
    \label{equ:DNLS_DST}
\end{equation}
With the procedure exhaustively discussed and showed in Sec.~\ref{sec:DNLS_full}, the continuity equation obtained from Eq.~\ref{equ:DNLS_sh} and Eq.~\ref{equ:DNLS_cc_sh} is still the same:
\begin{equation}
    i\frac{d}{dT}|u_{i}|^{2} - C\left(u_{i}\left[u_{i+1}^{*} + u_{i-1}^{*}\right] -\left[u_{i+1} + u_{i-1}\right]u_{i}^{*} \right)=0.
    \label{equ:conserve_sh}
\end{equation}
but Eq.~\ref{equ:tweve} slightly modifies (as we used multiplication by $(u_{i+1}^{*}+u_{i-1}^{*})$ and $-(u_{i+1}+u_{i-1})$ instead of $(u_{i+1}^{*}+u_{i-1}^{*}-2u_{i}^{*})$ and $-(u_{i+1}+u_{i-1}-2u_{i})$):
\begin{equation}
\begin{split}
    &i\left(\frac{d u_{i}}{d T} \left[u_{i+1}^{*} + u_{i-1}^{*} \right] + \left[u_{i+1} + u_{i-1} \right]\frac{d u_{i}^{*}}{d T}\right) + \\
    & +\gamma|u_{0/i}|^{2}\left(u_{i}\left[u_{i+1}^{*} + u_{i-1}^{*}\right] - \left[u_{i+1} + u_{i-1}\right]u_{i}^{*}\right) = 0
\end{split}
\end{equation}
yielding the basic Hamiltonian $H_{B}$:
\begin{equation}
    \frac{d H_{B}}{d T}+C\left(\frac{d u_{i}}{d T} \left[u_{i+1}^{*} + u_{i-1}^{*} \right] + \left[u_{i+1} + u_{i-1} \right]\frac{d u_{i}^{*}}{d T}\right)=0 
\end{equation}
with 
\begin{equation}
    H_{B}:=\gamma|u_{0/i}|^{2}|u_{i}|^{2}.
\end{equation}
As compared to Eq.~\ref{equ:sixteen}, it does not have a $2C-$term. This $2C-$term got lost, as we chose strictly next-neighbour coupling instead of diffractive coupling.

Under consideration of optical losses and cavity detuning, we get:
\begin{equation}
\begin{split}
    i(\gamma|u_{0/i}|^{2}&-\Delta)    
    \left(\frac{d}{d T}|u_{i}|^2 + \alpha|u_{i}|^{2}\right) +\\
    &+ i C\left(\frac{d u_{i}}{d T}\left[u_{i+1}^{*} + u_{i-1}^{*} \right] + \left[u_{i+1} + u_{i-1} \right]\frac{d u_{i}^{*}}{d T}\right) + \\
     & + iC\frac{\alpha}{2}\left(u_{i} \left[u_{i+1}^{*} + u_{i-1}^{*} \right] + \left[u_{i+1} + u_{i-1}\right]u_{i}^{*}\right)=0.
\end{split}
\label{equ:loss_cons}
\end{equation}
Suppose, $C=0$ in Eq.~\ref{equ:loss_cons}. We can use the plane-wave ansatz $u_{n} = u_{0}e^{i(\omega T + \kappa n)} =: \hat{u}_{i}e^{i\kappa n}$ to solve this equation:
\begin{equation}
    \frac{d}{d T}|\hat{u}_{i}|^{2} e^{\alpha T} + \alpha|\hat{u}_{i}|^{2}e^{\alpha T} = 0
    \label{equ:loss_multiply1}
\end{equation}
where we multiplied the expression by $e^{\alpha T}$ for the sake of convenience. Now, we see that this expression looks very much like the product rule of integration:
\begin{equation}
    \frac{d}{d T}|\hat{u}_{i}|^{2} e^{\alpha T} + \alpha|\hat{u}_{i}|^{2}e^{\alpha T} =\frac{d}{dT}\left(|\hat{u}_{i}|^{2} e^{\alpha T}\right)= 0
\end{equation}
which allows us to include the real-valued factor $e^{\alpha T}$ in the  plane-wave ansatz:  
\begin{equation}
    u_{n} = u_{0}e^{i(\omega T + \kappa n)}e^{\frac{\alpha}{2}T}.
\end{equation}
For $\alpha<0,$ $e^{\frac{\alpha}{2}T}$ denotes the exponential decay of the amplitude in the resonator $n$ due to optical losses. 

With consideration, we made in Eq.~\ref{equ:coupling_term} and having multiplied the coupling terms by $e^{\alpha T},$ we see that they also can be integrated delivering the Ising Hamiltonian: 
\begin{equation}
    \bar{H}_{I} = \left(\gamma|u_{0/i}|^{2}-\Delta\right)|u_{i}|^{2} +C[u_{i+1} + u_{i-1}]u_{i}^{*} = \bar{E}e^{\frac{\alpha}{2}T} \neq const.
    \label{equ:Ising_time}
\end{equation}
Due to optical losses, the Hamiltonian is not time-independent any more, i.e. $\bar{H}_{I}=\bar{H}_{I}(T)$ which perfectly makes sense as the system is not closed any more.

Taking into account the pump, we get:
\begin{equation}
\begin{split}
    i(\gamma|u_{0/i}|^{2}&-\Delta)    
    \left(\frac{d}{d T}|u_{i}|^2 + \alpha|u_{i}|^{2} - \mu_{i}\left(u_{i}u + u^{*}u_{i}^{*}\right)\right) +\\
    &+ i C\left(\frac{d u_{i}}{d T}\left[u_{i+1}^{*} + u_{i-1}^{*} \right] + \left[u_{i+1} + u_{i-1} \right]\frac{d u_{i}^{*}}{d T}\right) + \\
     & + i\frac{\alpha}{2} C\left(u_{i} \left[u_{i+1}^{*} + u_{i-1}^{*} \right] + \left[u_{i+1} + u_{i-1}\right]u_{i}^{*}\right)-\\
     & - i\mu_{i} C\left(u^{*} \left[u_{i+1}^{*} + u_{i-1}^{*} \right] + \left[u_{i+1} + u_{i-1}\right]u\right) =0.
\end{split}
\label{equ:Ham_fast}
\end{equation}
The last term describes how the pump couples to the neighbouring resonators. We can neglect it under the assumption of weak coupling and small pump amplitude. So, we need to integrate the following equation:
\begin{equation}
\frac{d\bar{H}_{I}}{dT}-i\mu_{i}(\gamma|u_{0/i}|^{2}-\Delta)
     (u_{i}u + u^{*}u_{i}^{*})=0.
\end{equation}

Let's have a closer look at the coupling between the resonator field $u_{n}$ and the pump $u$ (again, I use the index $n$ instead of $i$ to avoid confusion). Let's again use the plane-wave ansatz, i.e. $u_{n}=u_{0/n}\exp{(i(\omega T + \kappa n))}\exp{(\frac{\alpha}{2}T})$ with frequency $\omega$ that is provided by the pump $u=u_{p}\exp{(i\omega T)}$ and wave number $\kappa.$ In this case, we get:
\begin{equation}
   \mu_{i}[u_{n}u + u^{*}u_{n}^{*}] = 2\mu_{i}u_{0/n}u_{p}\cos{(2\omega T + \kappa n)}\exp{(\alpha T)}
\end{equation}
where we multiplied the expression by $\exp{(\alpha T)}$ as we did it in Eq.~\ref{equ:loss_multiply1}.

This expression shows that the coupling between the resonator field and the pump generates a field  that oscillates at twice the frequency. Integration over $T$ yields:
\begin{equation}
   \begin{split}
&2\mu_{i}u_{0/n}u_{p}\int\cos{(2\omega T + \kappa n)}\exp{\alpha T})dT=\\
   &=\frac{2\mu_{i}u_{0/n}u_{p}}{\alpha^{2}+ (2\omega)^{2}}\left(\alpha 2\cos{(2\omega T + \kappa n)} + 2\cdot 2\omega\sin{(2\omega T + \kappa n)}\right)\exp{(\alpha T)} + \\
   &+const.
    \end{split}
    \label{equ:coupl_int}
\end{equation}
This function has three minima and serves as a potential. Two pronounced minima are located at $\pm 2\omega.$ They provide potential wells for phase solitons as reported in Ref. [1]. The third minimum is at $\omega = 0,$ its depth depends on the value of $\alpha,$ $\alpha<0:$ the bigger the value of $|\alpha|,$ the deeper the well it creates. From the physical point of view, the minimum at $\omega = 0$ has no meaning. However, it competes with two other potential minima effectively preventing phase solitons at $\pm 2\omega$ from appearing if the system chooses to occupy the well at $\omega = 0.$ 

Also, we see that the whole expression is divided by $(\alpha^{2}+ (2\omega)^{2})$ which, in physical terms, is a huge value. This means that the amplitude $\frac{2\mu_{i}u_{0/n}u_{p}}{\alpha^{2}+ (2\omega)^{2}}$ is very small, which in turn means that the coupling between the pump and the resonator field applies only a very small modulation with frequency $2\omega$ to the resonator field. In the normal-dispersion regime, this modulation will have no impact. In the anomalous regime, due to modulational instability (MI), it will lead to the break-up of a continuous field into pulses (solitons). In fact, MI is one of the possible regimes observed in the experimental setup reported in Ref. [1]. 

Transforming Eq.~\ref{equ:coupl_int} back to the multiplicative form between the fields, we get
\begin{equation}
    \frac{2\mu_{i}}{\alpha^{2}+ (2\omega)^{2}}\left(\alpha[u_{n}u + u^{*}u^{*}_{i}]-2i\omega[u_{n}u-u^{*}u^{*}_{i}]\right) + const = :P(u_{n}, u).
\end{equation}
The Ising Hamiltonian updates now to
\begin{equation}
    \bar{H}_{I} = \left(\gamma|u_{0/i}|^{2}-\Delta\right)|u_{i}|^{2} +C[u_{i+1} + u_{i-1}]u_{i}^{*} - P(u_{i}, u).
    \label{equ:Ising_time_pump}
\end{equation}
Generalisation to a DST equation (cf. Eq.~\ref{equ:DNLS_DST}) as well as dropping the assumption that the optical Kerr effect depends only on the initial amplitude leads to the final version of the Ising Hamiltonian for $i-$th resonator and coupling to all remaining $N-1$ resonators:
\begin{equation}
\begin{split}
    {H}_{Ising} &= \gamma|u_{i}|^{4}-\Delta|u_{i}|^{2} +C\left(\sum_{k\neq i}^{N}u_{k}\right)u_{i}^{*} - P(u_{i}, u),\\
    P(u_{i}, u) &= \alpha P_{0/i}\left([u_{n}u + u^{*}u^{*}_{i}]-i\frac{2\omega}{\alpha}[u_{n}u-u^{*}u^{*}_{i}]\right),\quad P_{0/i}\ll 1.  
  \end{split}
  \label{equ:Ising_final}
\end{equation}
Comparison with Ref. [7, 8] as well as the classical Ising model dictates us the values of the parameters. Thus, in Eq.~\ref{equ:Ising_final}, the optical losses $\alpha$ are negative, the coupling parameter $C$ is negative, the cavity detuning can be both, positive and negative. However, for the existence of solitons, it has to have the opposite sign to the positive (Kerr) nonlinear parameter, i.e. to be negative.     

\section{A Note on the Dispersion}
With the plane-wave ansatz, the contribution of the dispersive term $\beta_{2}\frac{\partial^{2}}{\partial\tau^{2}}u\rightarrow -\beta_{2}\omega^{2}u$ cancels out (disappears) when the continuity equation of the form Eq.~\ref{equ:conserve_sh} is calculated. But it appears when Eq.~\ref{equ:DNLS_sh} is multiplied by $(u_{i+1}^{*}+u_{i-1}^{*}),$ Eq.~\ref{equ:DNLS_cc_sh} by $-(u_{i+1}+u_{i-1}),$ and the resulting expressions are added:
\begin{equation}
\begin{split}
    &i\left(\frac{u_{i}}{dT}\left(u^{*}_{i+1} + u^{*}_{i-1}\right) + \left(u_{i+1} + u_{i-1}\right)\frac{d u^{*}}{dT}\right)-\\
    &-\beta_{2}\omega^{2}\left(u_{i}[u_{i+1}^{*} + u_{i-1}^{*}]- [u_{i+1} + u_{i-1}]u_{i}^{*}  \right) = 0.
\end{split}
\end{equation}
To get rid of this term, we need to multiply the following expression by $(\gamma|u_{0/i}|^{2} - \beta_{2}\omega^{2} - \Delta)$ rather than by $(\gamma|u_{0/i}|^{2} - \Delta).$ In the anomalous dispersion regime, the relation $(\gamma|u_{0/i}|^{2} - \beta_{2}\omega^{2})$ governs the evolution of optical solitons, something we in fact need to be able to observe phase solitons as described in Ref. [1]. The effect of the dispersion can be integrated into the effective cavity detuning, $\Delta_{eff} = (\beta_{2}\omega^{2} + \Delta)$ to create a dispersion-dependent Ising Hamiltonian:
\begin{equation}
    {H}_{Ising} = \gamma|u_{i}|^{4}-\Delta_{eff}|u_{i}|^{2} +C\left(\sum_{k\neq i}^{N}u_{k}\right)u_{i}^{*} - P(u_{i}, u)
  \label{equ:Ising_final_disp}
\end{equation}
Again, $\gamma \geq 0,$ $\Delta_{eff}$ can be positive or negative, or $\Delta_{eff}=0,$ $C <0,$ $P(u, u)<0.$

\section{References}
\begin{enumerate}

\item N. Englebert et al., \emph{Parametrically driven Kerr cavity solitons.} Nature Photonics, 15, pp. 857–861 (2021)
\item M. Zajnulina, \emph{Nonlinear Phenomena in Coupled Silicon-on-Insulator Waveguides.} Diploma thesis, TU Berlin (2012)
\item N. A. Kudryashov, \emph{Hamiltonians of the Generalized Nonlinear Schr\"odinger Equations.} Mathematics 11, 2304 (2023) 
\item V. I. Nekorkin \& M. G. Velarde, \emph{Synergetic Phenomena in Active Lattices. Patterns, Waves, Solitons, Chaos.} Springer (2002)
\item J. Ch. Eilbeck \& M. Johansson, \emph{The Discrete Nonlinear Schr\"odinger Equation - 20 Years On.} Localization and Energy Transfer in Nonlinear Systems, pp. 44-67 (2003).

\item R. Hamerly et al., \emph{Experimental investigation of performance differences between Coherent Ising Machines and a quantum annealer.} Science Advances 5(5) (2019)
\item H. Goto et al., \emph{Combinatorial optimization by simulating adiabatic bifurcations in nonlinear Hamiltonian systems.} Science Advances 5(4) (2019)
\item F. B\"{o}hm et al., \emph{A poor man's coherent Ising machine based on opto-electronic feedback system for solving optimization problems.} Nature Communications 10:3538 (2019) 
\end{enumerate}

\end{document}